\documentclass{article}
\usepackage{spconf,amsmath,graphicx,url,comment,tipa,multirow}
\usepackage[backend=biber, style=ieee, citestyle=numeric, maxnames=3]{biblatex}
\usepackage[fontsize=9.25pt]{fontsize}
\usepackage{xcolor}
\usepackage{kotex}
\newcommand{\jn}[1]{\textcolor{magenta}{(JN: #1)}}
\newcommand{\sbc}[1]{\textcolor{orange}{(SBC: #1)}}

\title{A Melody-Unsupervision Model for Singing Voice Synthesis}

\name{Soonbeom Choi, Juhan Nam}
\address{Graduate School of Culture Technology, KAIST, Daejeon, South Korea}
\begin{document}
%
\maketitle
\begin{abstract} 
Recent studies in singing voice synthesis have achieved high-quality results leveraging advances in text-to-speech models based on deep neural networks. One of the main issues in training singing voice synthesis models is that they require melody and lyric labels to be temporally aligned with audio data. The temporal alignment is a time-exhausting manual work in preparing for the training data. To address the issue, we propose a melody-unsupervision model that requires only audio-and-lyrics pairs without temporal alignment in training time but generates singing voice audio given a melody and lyrics input in inference time. The proposed model is composed of a phoneme classifier and a singing voice generator jointly trained in an end-to-end manner. The model can be fine-tuned by adjusting the amount of supervision with temporally aligned melody labels. Through experiments in melody-unsupervision and semi-supervision settings, we compare the audio quality of synthesized singing voice. We also show that the proposed model is capable of being trained with speech audio and text labels but can generate singing voice in inference time.  

\end{abstract}
\section{Introduction}
\label{sec:introduction} 
Singing voice synthesis (SVS) is a task that generates natural vocal sounds from musical notes (melody) and time-aligned text (lyrics). Leveraging the advances in neural text-to-speech (TTS) models \cite{Tacotron2017, Tacotron2_2018, DeepVoice3_2018, Tachibana2018EfficientlyTT} and neural vocoders \cite{WaveNet2016, Waveglow2019, ParallelWaveGAN2020, HiFiGAN2020}, SVS models based on deep neural network have made great improvements in terms of sound quality and naturalness. Many of them have been implemented with an autoregressive model \cite{NPSS2017, Kim2018, Nakamura2019, Lee2019, BEGANSing2020, DeepSinger2020}. Recent SVS models based on transformer have shifted the paradigm to a feed-forward manner, overcoming the limitations of autoregressive models such as slow speed and error propagation in inference time \cite{Blaauw2020, XiaoiceSing2020, NSinger2021}.



Unlike TTS, SVS takes melody as an additional input source and syllables in lyrics must be aligned with notes in the melody. Furthermore, the onset and duration of notes should be temporally aligned with audio data in frame-level for training the SVS models \cite{Lee2019,BEGANSing2020,NSinger2021}. Since the frame-level labeling is a highly time-consuming labor, some SVS models tackled the issue using a duration predictor in the model \cite{Blaauw2020, XiaoiceSing2020}, a separate audio-to-lyrics alignment model \cite{DeepSinger2020} or time-aligned acoustic features \cite{DurIANSC2020}. However, the duration prediction methods still require duration labels and the separate audio-to-lyrics alignment makes the data processing complicated. Training with time-aligned acoustic features in \cite{DurIANSC2020} is considered as a voice conversion problem which needs a reference voice in inference time.



To address this issue, we propose a melody-unsupervision model for SVS which requires only audio-and-lyrics pairs without their temporal alignment in the training time and therefore significantly reduce the labeling efforts. The model is composed of a phoneme classifier and a singing voice generator, which are trained jointly in an end-to-end manner. The phoneme classifier is trained with the lyrics text using the connectionist temporal classification (CTC) loss and a unvoice penalty term. The singing voice generator is trained using the predicted phonemes and fundamental frequency (F0) pseudo labels with the audio data. This model can be also trained with melody supervision using only the singing voice generator without any change in model configuration. 
We show that the SVS model trained with melody-unsupervision is capable of generating meaningful vocal sounds. We also fine-tuned the model by adjusting the amount of supervision with melody and temporally alignment. We show that the gradual fine-tuning with melody-supervision progressively increases the performance. Finally, we show that the proposed model can be trained even with TTS data (speech audio and text labels) but can generate singing voice in inference time.  



\begin{figure*}[!t]
\centering
\vspace{-2mm}
\includegraphics[width=16.5cm, clip]{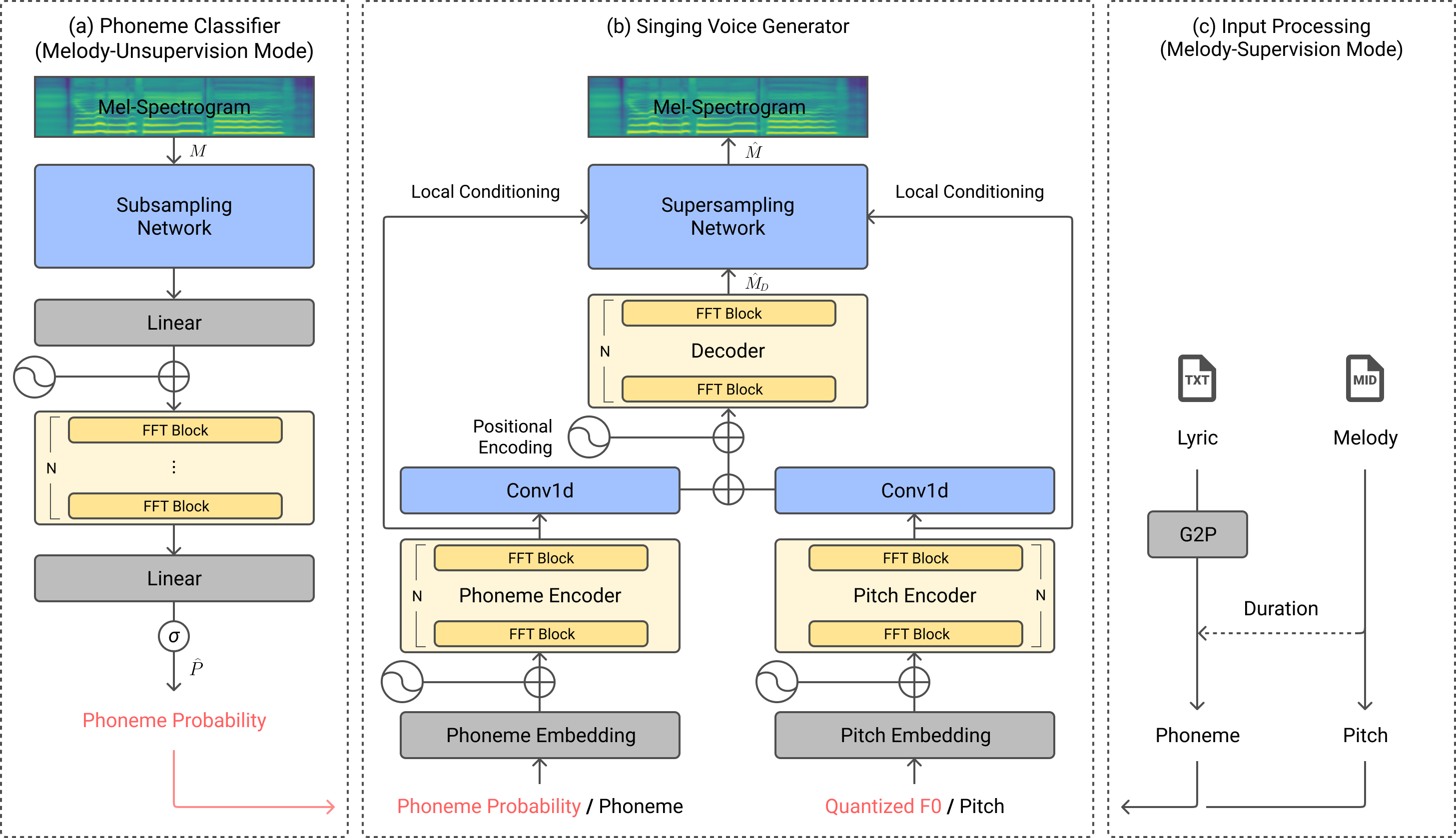}
\vspace{-2mm}
\caption{Overview of the melody-unsupervision model for singing voice synthesis. The singing voice generator (b) is trained with phonetic information from the phoneme classifier (a) and melodic information from quantized F0 in melody-unsupervision mode (red). The phoneme and pitch inputs are replaced with the time-aligned lyric and melody labels (c) in melody-supervision mode (black).}
\label{fig:model}
\end{figure*}

\section{Method}
\label{sec:proposed_method}

The proposed model consists of a phoneme classifier and singing voice generator as shown in \figurename~\ref{fig:model}. In melody-unsupervision mode, the phoneme classifier takes mel-spectrogram as input and predicts a phoneme probability. 
The singing voice generator takes the predicted phoneme probability from the phoneme classifier and quantized pseudo pitch labels from a pitch estimator as input, and predicts mel-spectrogram. The phoneme classifier and the singing voice generator are jointly trained. In melody-supervision or inference mode, the input of the singing voice generator is replaced by a given lyrics and musical score. 
Finally, a neural vocoder generates the output waveforms conditioned on mel-spectrogram from the singing voice generator. The details of each module are explained below.  


\subsection{Phoneme Classifier}
We approach phoneme classification as automatic speech recognition (ASR) in pronunciation level. Phoneme prediction is relatively simple compared to word prediction which requires a large vocabulary size and more complex temporal dependency. We built the phoneme classifier by simplifying non-autoregressive ASR models \cite{TransformerASR2018, Nakatani2019, Salazar2019, Miao2020}. Specifically, we adopted a convolution subsampling network from Conformer \cite{Conformer2020} to down-sample the mel-spectrogram input ($M$) and used a transformer encoder to make a prediction as shown on the left side of \figurename~\ref{fig:model}. 

\subsection{Singing Voice Generator}
\subsubsection{Encoder-Decoder}
The encoder and decoder architectures in the singing voice generator are based on transformer blocks as shown on the middle of \figurename~\ref{fig:model}. The transformer block is composed of multi-head self-attention and position-wise fully-connected feed-forward network, and operates in non-autoregressive manner. It is similar to the non-autoregressive transformer architectures in \cite{DeepSinger2020, NSinger2021}. We used two transformer encoders with an identical architecture for phoneme input and pitch input, respectively. Each of them has its own embedding layer and positional encoding. The transformer decoder takes the sum of phoneme and pitch encoders and generates the down-sampled mel-spectrogram ($\hat{M_D}$). 

\subsubsection{Supersampling Network}
Since the decoder generates down-sampled mel-spectrogram, we added a supersampling network to restore the frame rate of mel-spectrogram. While this supersampling network is similar to the super-resolution network in \cite{Lee2019, Lee2020}, it upscales mel-spectrogram only time-wise. The supersampling network has stacks of highway blocks with transposed convolution networks and uses local conditioning from the phoneme encoder output and pitch encoder output as in \cite{Lee2019, Lee2020}. 

\subsection{Neural Vocoder}
We use HiFi-GAN as a neural vocoder. HiFi-GAN is conditioned on mel-spectrogram and can synthesize high-fidelity voice faster than real-time \cite{HiFiGAN2020}. HiFi-GAN achieves the high performance by applying multi-receptive field fusion on generator which sums outputs from multiple residual blocks with different kernel sizes and dilation sizes, multi-period discriminator and multi-scale discriminator. We trained the HiFi-GAN module from scratch using our singing voice audio data. 


\begin{table*}[!t]
\centering
\resizebox{\textwidth}{!}{
\begin{tabular}{c|c|ccc|c|ccc|c|c} 
\hline
\multicolumn{2}{c|}{}                                 & \multicolumn{3}{c|}{\textbf{Supervised }}           & \textbf{Unsupervised } & \multicolumn{3}{c|}{\textbf{Semi-Supervised}}                 & \textbf{ATK }   & \textbf{GT}      \\ 
\hline
\multicolumn{2}{c|}{\# of songs with melody labels}   & 11              & 22              & 44              & 0                      & 11              & 22              & 44                        & 44              &    -               \\ 
\hline
\multirow{3}{*}{EN} & Pronunciation Acc               & 2.49 $\pm$ 0.32 & 2.85 $\pm$ 0.38 & 3.19 $\pm$ 0.45 & 2.35 $\pm$ 0.17        & 2.95 $\pm$ 0.32 & 3.12 $\pm$ 0.33 & \textbf{3.31~$\pm$ 0.51}  & 2.31 $\pm$ 0.51 & 4.89 $\pm$ 0.07  \\
                    & Naturalness                     & 2.82 $\pm$ 0.28 & 3.33 $\pm$ 0.29 & 3.36 $\pm$ 0.17 & 2.73 $\pm$ 0.23        & 3.21 $\pm$ 0.27 & 3.29 $\pm$ 0.17 & \textbf{3.56 $\pm$ 0.33 } & 2.83 $\pm$ 0.18 & 4.87 $\pm$ 0.09  \\
                    & Overall Quality                 & 2.61 $\pm$ 0.27 & 3.14 $\pm$ 0.37 & 3.27 $\pm$ 0.29 & 2.58 $\pm$ 0.21        & 3.14 $\pm$ 0.23 & 3.20 $\pm$ 0.22 & \textbf{3.43 $\pm$ 0.42 } & 2.58 $\pm$ 0.37 & 4.89 $\pm$ 0.08  \\ 
\hline
\multirow{3}{*}{KR} & Pronunciation Acc               & 2.63 $\pm$ 0.44 & 3.14 $\pm$ 0.40 & 3.40 $\pm$ 0.43 & 2.82 $\pm$ 0.44        & 3.33 $\pm$ 0.62 & 3.47 $\pm$ 0.27 & \textbf{3.53 $\pm$ 0.38 } & 2.93 $\pm$ 0.34 & 4.77 $\pm$ 0.11  \\
                    & Naturalness                     & 2.98 $\pm$ 0.24 & 3.28 $\pm$ 0.32 & 3.43 $\pm$ 0.16 & 2.83 $\pm$ 0.27        & 3.44 $\pm$ 0.34 & 3.51 $\pm$ 0.09 & \textbf{3.52 $\pm$ 0.16 } & 3.34 $\pm$ 0.35 & 4.80 $\pm$ 0.12  \\
                    & Overall Quality                 & 2.79 $\pm$ 0.34 & 3.16 $\pm$ 0.35 & 3.41 $\pm$ 0.18 & 2.79 $\pm$ 0.35        & 3.33 $\pm$ 0.44 & 3.41 $\pm$ 0.18 & \textbf{3.50 $\pm$ 0.27 } & 3.09 $\pm$ 0.40 & 4.87 $\pm$ 0.09  \\
\hline
\end{tabular}
}
\caption{Qualitative evaluation results in MOS for both English (EN) and Korean (KR)}
\label{table:mos}
\end{table*}

\vspace{-2mm}
\subsection{Melody-Unsupervision Mode}
In melody unsupervision mode, the phoneme classifier predicts the phoneme probabilities in each frame. Since phonetic timing information is not provided by the ground truth, we used the CTC loss. To predict accurate phonetic timing, the model needs to predict blank only for silence or unvoiced frames. We observed that the model trained with the CTC loss tends to predict blanks for sustained voice frames in our experience. 
Thus, we added an unvoice penalty term to suppress such blank predictions on voiced frames. We estimated the unvoice state using the Harvest F0 estimation algorithm \cite{Harvest2017}, and calculated the unvoice predictions for voiced frames as the penalty. The singing voice generator is trained with the L1 loss for both the down-sampled mel-spectrogram ($\hat{M_{D}}$) and the super-sampled mel-spectrogram ($\hat{M}$). The following equations specify the entire loss of the proposed model:
\begin{equation}
\begin{split}
& \mathcal{L}_{C} = \mathcal{L}_{CTC}(log(\hat{P}), P) + \mathcal{L}_{UV} \\
& \mathcal{L}_{G} = \mathcal{L}_{1}(\hat{M_D}, M_D) + \mathcal{L}_{1}(\hat{M}, M) \\
& \mathcal{L} = \mathcal{L}_{C} + \mathcal{L}_{G}
\end{split}
\label{loss}
\end{equation}
where $P$ is the ground truth phonemes, $L_{UV}$ is the unvoice penalty, $L_C$ is the phoneme classifier loss and $L_G$ is the singing voice generator loss.

We calculated the input of the phoneme encoder by multiplying the predicted phoneme probabilities with the phoneme embedding matrix instead of taking a single phoneme with the highest probability. This makes the gradient from the singing voice generator flow to the phoneme classifier so that the modules are trained in an end-to-end manner.
For the pitch encoder, we used F0 estimated from Harvest. We quantized the F0 to semi-tone steps in MIDI note scale. The quantized F0 can be considered as a pseudo label for melody input from musical scores.

\subsection{Melody Supervision Mode}
In melody supervision mode, only the singing voice generator is trained using the phoneme sequence from lyrics and the pitch sequence from musical score like conventional SVS systems.

\section{Experiments}
\label{sec:experiments}

\vspace{-1mm}
\subsection{Dataset}
Most of the previous SVS systems were evaluated using a private dataset which makes the results unreproducible. We trained our proposed model mainly using the children's song dataset (CSD) which is a public dataset composed of 50 Korean and 50 English children songs with melody transcriptions and lyrics. We trained the model using two languages separately. We used 44 songs for training the remaining 6 songs for inference.

\subsection{Input Processing}
We extracted phonemes using grapheme-to-phoneme algorithms \cite{KoreanG2p, EnglishG2p} for both languages. We represented both phonemes and notes as an one-hot vector and repeated the vectors for a given duration. For both Korean and English we assigned consonants to the beginning or the end frame of a note, and vowels to the remaining frames as in \cite{Kim2018, Blaauw2020}. While an average phoneme duration from a pre-trained alignment model was computed as a consonant length in \cite{Blaauw2020}, we fixed the length to 10 msec, which corresponds to two frames in our settings, for all consonants.

Each audio recording in CSD is a few minutes long so we split it into audio chunks by detecting silence or breath sounds between melodies. As a result, the input audio length ranged from 2.5 to 12 sec. We resampled the input audio to 24000 Hz and computed mel-spectrogram using 1024-point fast Fourier transform with a Hann window of 1024 samples and mel-filterbank with 80 bins. The mel-spectrogram was compressed in dB and normalized with the minimum value of -100dB and maximum value of 0dB.


\subsection{Training}
We trained the proposed SVS model in three different settings. The first setting is to train the model using the entire 44 songs solely in melody-unsupervision mode. The second setting is to fine-tune the model trained in melody-unsupervision mode (the first setting) with a partial set of melody and lyric labels in melody-supervision mode. We increased the partial set progressively from 5, 11, 22 to 44 songs for both languages. The last setting is to train the model using the partial set of training data solely in melody-supervision mode. In the inference phase, we synthesized the singing voice from given melody and lyrics with the input configuration in melody-supervision mode. 


The models were trained using the Adam optimizer \cite{Adam} with $\beta_{1} = 0.5, \beta_{2} = 0.9$ and $\epsilon = 10^{-8}$. We scheduled the learning rate using three cycles of cosine annealing with warm-ups. The initial learning rate was set to $2.5 \times 10^{-4}$ and every cycle the model was trained for 10,000 steps and the learning rate was reduced by half. In the semi-supervised setting, the cosine annealing restarts two times in melody-unsupervision mode and one time in the following melody-unsupervision mode. 

We compared our model to the adversarially trained end-to-end Korean SVS model (ATK) \cite{Lee2019} as a reference. For fair comparison in sound quality, we modified the super-resolution network in ATK to predict mel-spectrogram as in \cite{NSinger2021} and used HiFi-GAN to generate audio. Since the source code of the ATK model is not available, we re-implemented it on our own and trained it using CSD. 

\section{Results and Discussion}
\subsection{Qualitative Evaluation}
We conducted a listening test for qualitative evaluation. To evaluate the generated singing voice in different aspects, we used three evaluation criteria: pronunciation accuracy, naturalness, and overall quality \cite{Kim2018, BEGANSing2020}. We did not include the sound quality in the criteria because it heavily relies on the neural vocoder. In the listening test, we compared the results from the proposed models in the three different training settings and the ATK model along with the ground truth (GT). 20 graduate students working on speech or music research participated in the listening test and evaluated the test examples in blind with mean opinion score (MOS). 

Table \ref{table:mos} shows the summary of MOS with mean and standard deviation. The models trained in melody-unsupervision mode (\textbf{Unsupervised}) are generally low in the MOS mean. As the the model is fine-tuned with more melody supervision (\textbf{Semi-Supervised}), the performance progressively increases. In particular, when the supervision starts (the number of songs with melody labels is 11), the MOS mean dramatically increases from 2.58 to 3.14 (EN), and 2.79 to 3.33 (KR) in overall quality). With the entire 44 songs, the semi-supervised models achieve the best MOS. This trend is observed for both English and Korean songs. The models trained in melody-supervision mode from scratch (\textbf{Supervised}) also show increasing performance as the number of songs with melody labels goes up. However, the overall MOS means of the supervised models are lower than those of the semi-supervised models for the same number of songs with melody labels. When the number of songs is not sufficient (11), the MOS means of the supervised models is even comparable to those of the unsupervised models. The ATK model, which was trained with the entire 44 songs in a supervised manner, shows generally lower MOS means than the propose models supervised or semi-supervised with the same number of songs. In particular, the gap is larger for English songs because the ATK model was not tuned for English songs \cite{Lee2019}.   


\begin{figure}[!t]
\vspace{-3mm}
\includegraphics[width=4.3cm, clip]{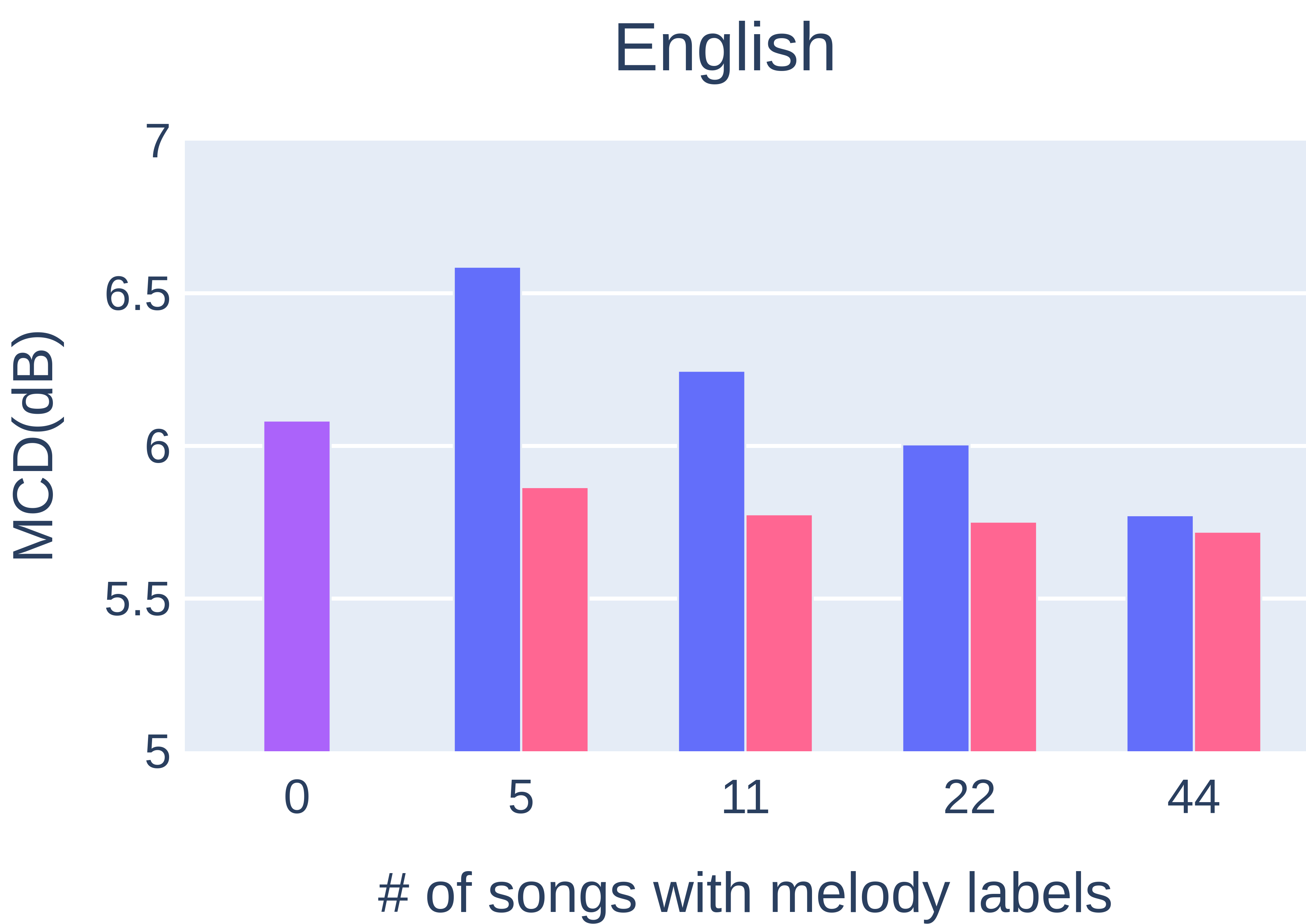}
\includegraphics[width=4.3cm, clip]{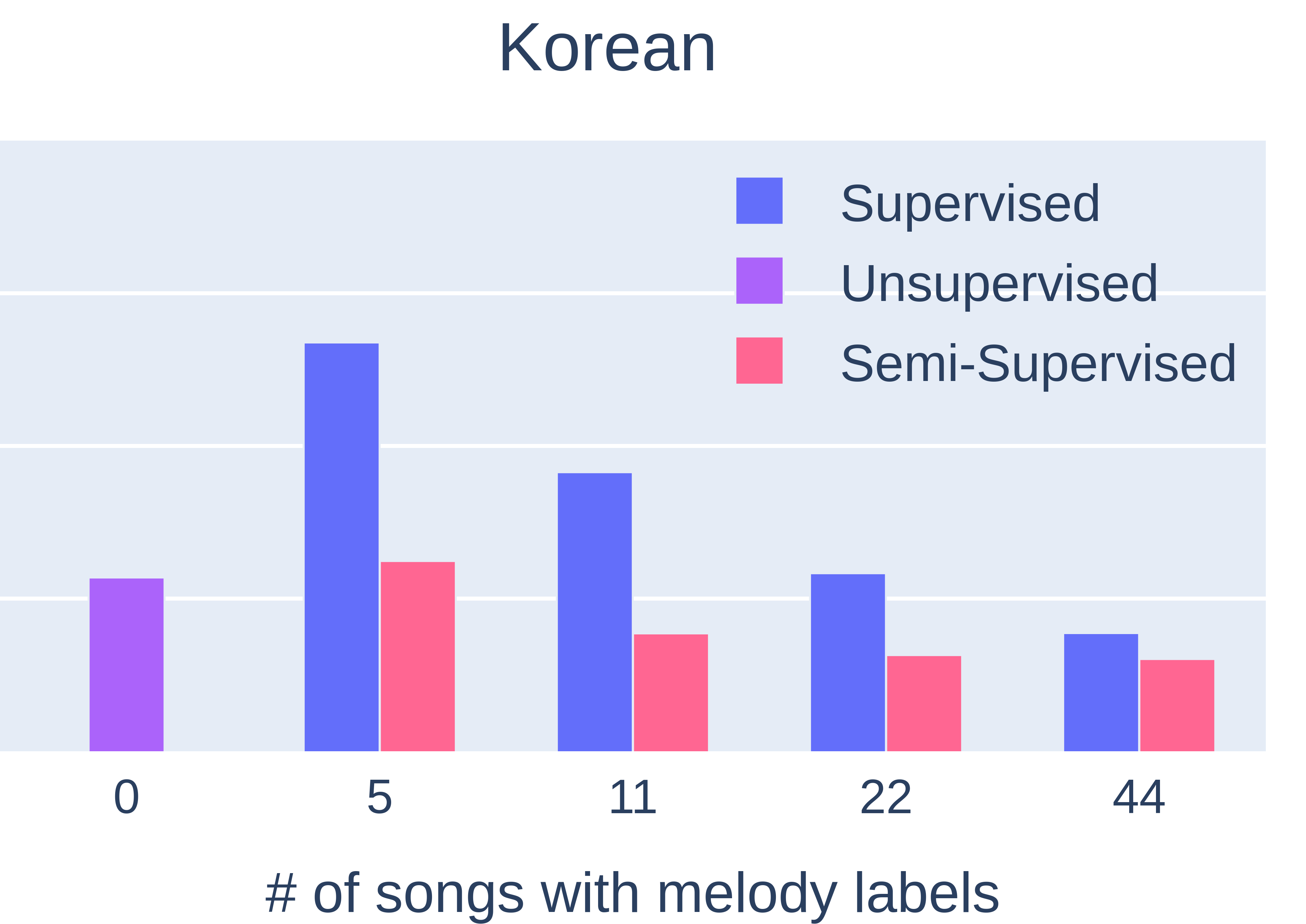}
\includegraphics[width=4.3cm, clip]{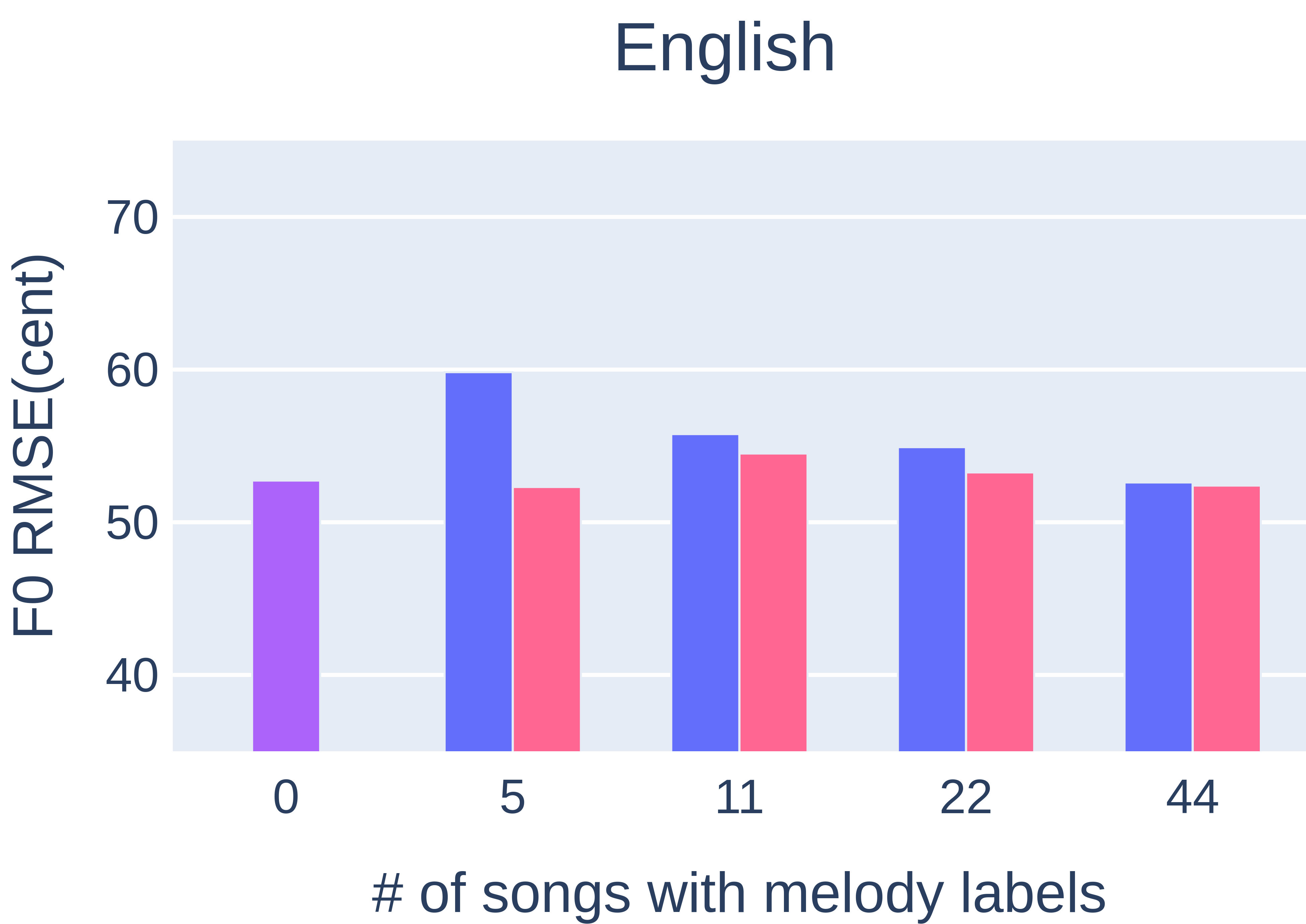}
\includegraphics[width=4.3cm, clip]{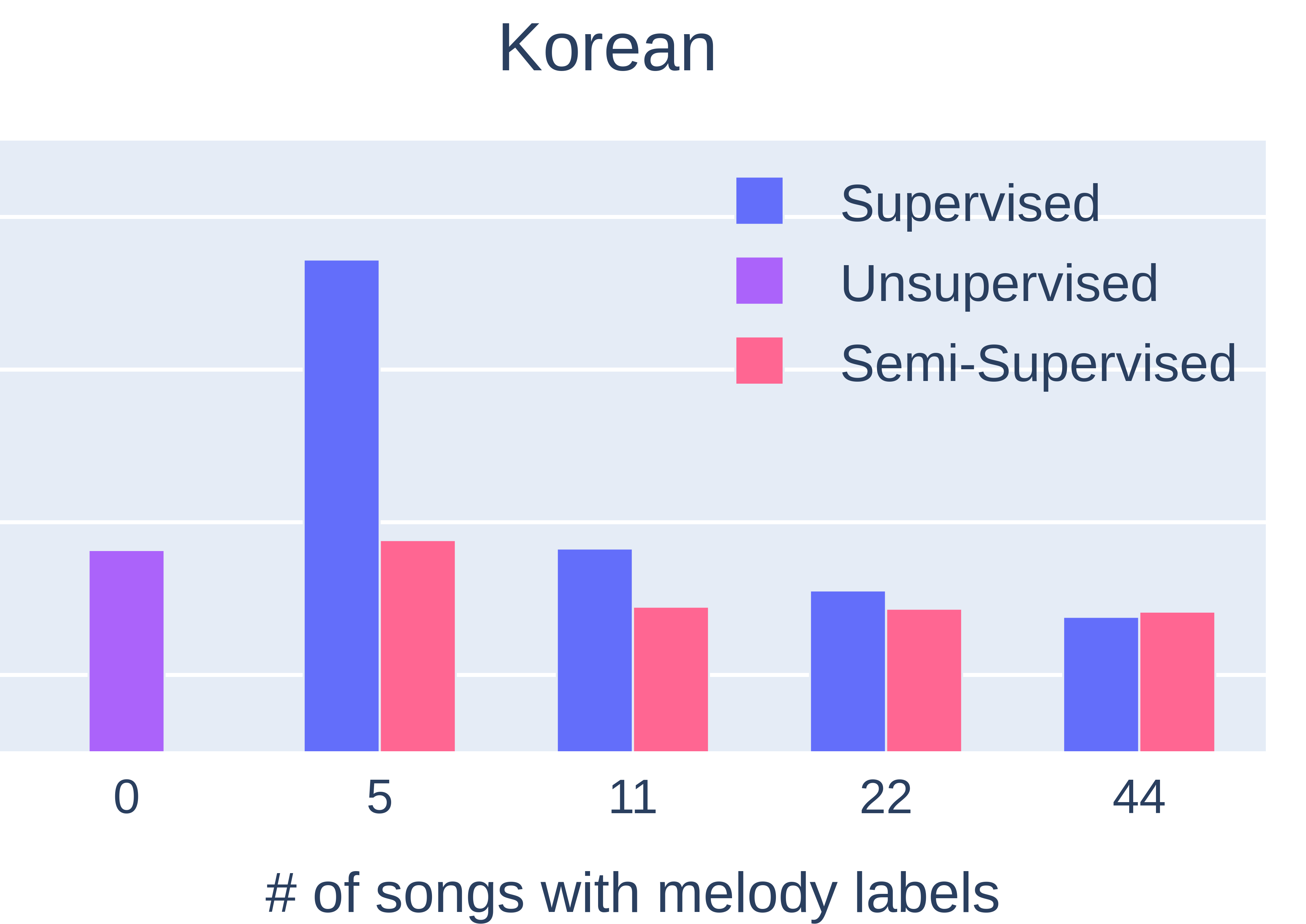}
\caption{MCD and F0 RMSE results with regards to the increasing number of songs with melody labels for the unsupervised, semi-supervised models, and supervised models}
\label{fig:chart_eval}
\end{figure}

\subsection{Quantitative Evaluation}
We adopted mel-cepstral distortion (MCD) and root mean square error (RMSE) of F0 for quantitative evaluation of the generated singing voice. 
\figurename~\ref{fig:chart_eval} shows the results with regards to the increasing number of songs with melody labels. The overall trends are similar to those in MOS but there are some differences. In MCD, when the number of songs with melody labels is 5 or 11, there are large gaps between the semi-supervised models and the supervised models, and the supervised models are worse in the unsupervised model in both languages. In F0 RMSE, the number of songs with melody labels does not affect the models much except the supervised model trained with 5 songs in Korean. This relatively small difference between the models is presumably because the quantized F0 in unsupervision mode is not always correct as pseudo labels.               


\vspace{-2mm}
\subsection{Phoneme Prediction}
\begin{figure}[!t]
\includegraphics[width=8.6cm, clip]{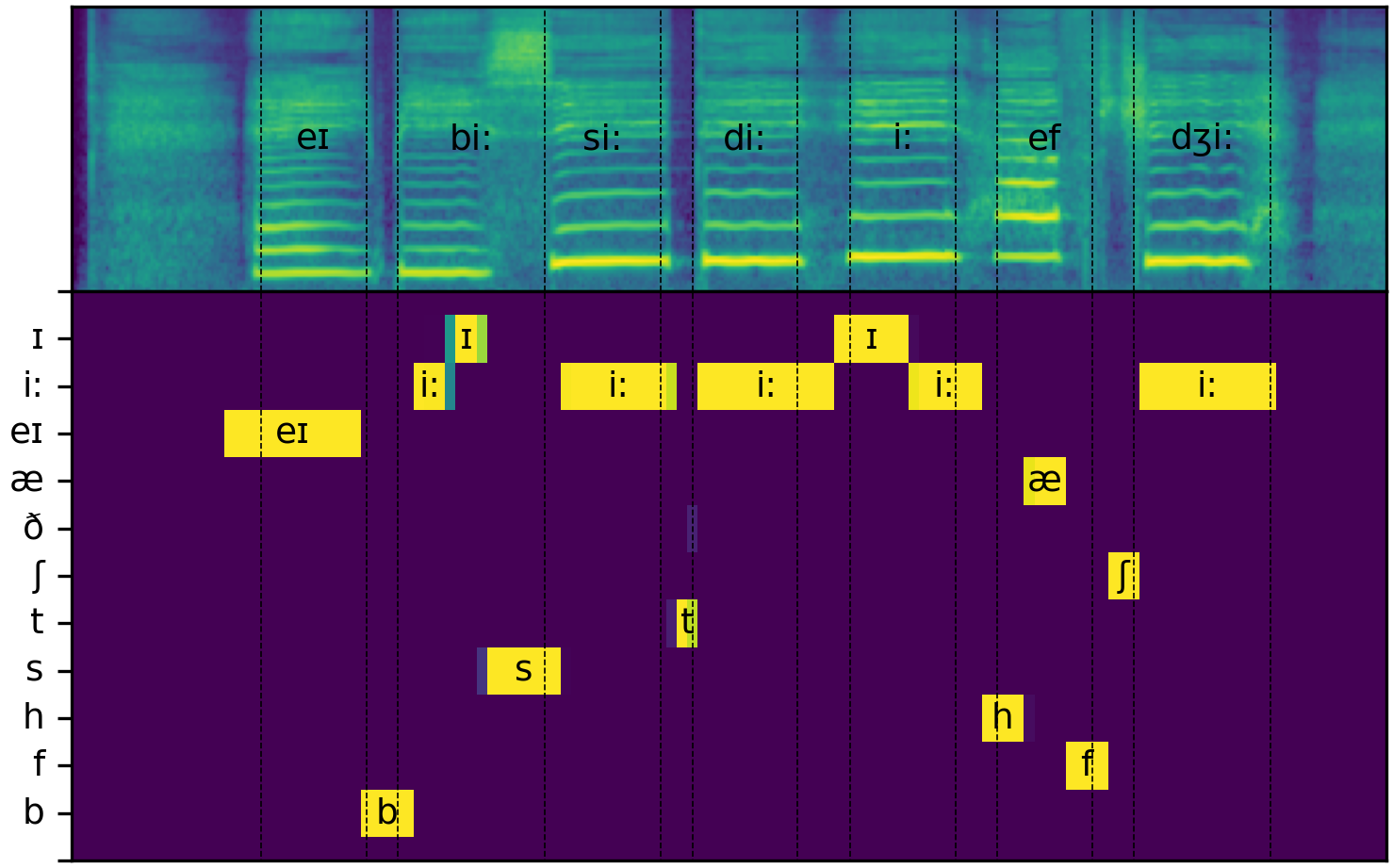}
\caption{Predicted phoneme probability (down) from a given mel-spectrogram overlayed with a ground truth phonemes (up). Phonemes are overlayed over probabilities when phoneme probabilities higher than 50\% are continued.}
\vspace{-2mm}
\label{fig:phoneme_prob_en}
\end{figure}

The proposed phoneme classifier can predict phoneme alignment without using phonetic timing labels using the CTC loss. The result significantly affects the pronunciation quality of the generated singing voice. As shown in \figurename~\ref{fig:phoneme_prob_en}, the model predicts distinguishable phoneme boundaries close to the ground truth and it has no alignment issue, although there is confusion between similar phonemes such as [t] and [d] or [\textesh] and [\textdyoghlig]. 
Although we could not measure the phoneme error rate due to the absence of frame-level phonetic timings, we instead measured the phoneme error rate within a syllable (PERS) using the note duration which counts wrong phoneme occurrences within a syllable. Overall PERS was 36.46\% for English and 26.13\% for Korean.

\subsection{Singing Voice Synthesis with TTS Data}
The proposed melody-supervision SVS model can be trained with audio and lyrics without the temporal alignment. This is the same condition as training TTS models and thus led us to train the proposed SVS model with LJSpeech, a well-known public speech dataset, to generate singing voice from speech data. From the results, we could observe that the model can reliably synthesize singing voices from speech data, being controlled with various melodic note sequences. The demo examples are available in the companion website. \footnote{All audio examples are available at \newline \url{https://soonbeomchoi.github.io/melody-unsupervised-blog/}}


\section{Conclusions}
We proposed a melody-unsupervision model for singing voice synthesis. It is composed of a phoneme classifier and a singing voice generator and they are trained jointly using audio and lyrics only. It also allows melody-supervision mode where only the singing voice generator is trained using audio, lyrics and melody with time-alignment. The results show that the unsupervised model generates singing voice with a reasonable quality and it can be improved further with a limited amount of melody labels, outperforming solely supervised models. This indicates a potential to scale up the training data easily for SVS. 


\vfill\pagebreak

\section{References}
\printbibliography[heading=none]

\end{document}